\documentclass{acm_proc_article-sp}
\usepackage{graphicx}

\begin{document}

\title{
Combining Semantic Wikis and\\
Controlled Natural Language
}

\numberofauthors{1}

\author{
  \alignauthor
  Tobias Kuhn\\
  \affaddr{Department of Informatics}\\
  \affaddr{University of Zurich, Switzerland}\\
  \email{tkuhn@ifi.uzh.ch}
}

\maketitle
\begin{abstract}
We demonstrate AceWiki that is a semantic wiki using the controlled natural language Attempto Controlled English (ACE). The goal is to enable easy creation and modification of ontologies through the web. Texts in ACE can automatically be translated into first-order logic and other languages, for example OWL. Previous evaluation showed that ordinary people are able to use AceWiki without being instructed.
\end{abstract}

\keywords{Controlled Natural Language, Attempto Controlled English (ACE), Semantic Web, Semantic Wiki, AceWiki, Ontology}

\section{Introduction}

Most of the tasks the Semantic Web is eventually supposed to fulfill rely on the availability of ontologies. However, the creation and maintenance of ontologies is difficult because a number of domain experts --- most of which are not familiar with formal languages --- have to agree on a conceptualization of the respective domain. For that reason, it is crucial for the future of the Semantic Web to provide tools that make the creation of ontologies easy for everybody.

AceWiki\footnote{See \cite{kuhn08acewiki}, \cite{kuhn08acewikionto}, and \texttt{http://attempto.ifi.uzh.ch/acewiki}} tackles this problem by combining semantic wikis with controlled natural language. The goal of AceWiki is to enable ordinary people with no background in formal languages to create expressive ontologies in a collaborative and intuitive way without the need of installing an application.

\section{Background}

There are several existing semantic wiki systems, see e.g. \cite{kuhn08acewiki} for a brief survey. Unfortunately, most of those wikis do not support expressive ontology languages in a general way. Furthermore, they are often hard to understand for people who are not familiar with the technical terms.

Attempto Controlled English (ACE)\footnote{See \cite{fuchs08summerschool} and \texttt{http://attempto.ifi.uzh.ch}} is the controlled natural language that is used for AceWiki. Being a subset of English, ACE looks completely natural. Restrictions of the syntax and the definition of a small set of interpretation rules make it a formal language that is automatically translatable into first-order logic. ACE covers a large part of natural English: singular and plural noun phrases, active and passive voice, relative phrases, anaphoric references, existential and universal quantifiers, negation, and much more.

ACE has been used as a natural language front-end to OWL with a bidirectional mapping of ACE to OWL \cite{kaljurand07phd}. AceWiki uses this for translating ACE sentences into OWL. The same work also introduces a Prot\'eg\'e plugin called ``ACE View'' which enables to manage ontologies in ACE within the Prot\'eg\'e environment.

\section{System}

\begin{figure}[tb]
  \begin{center}
    \includegraphics[width=8.4cm]{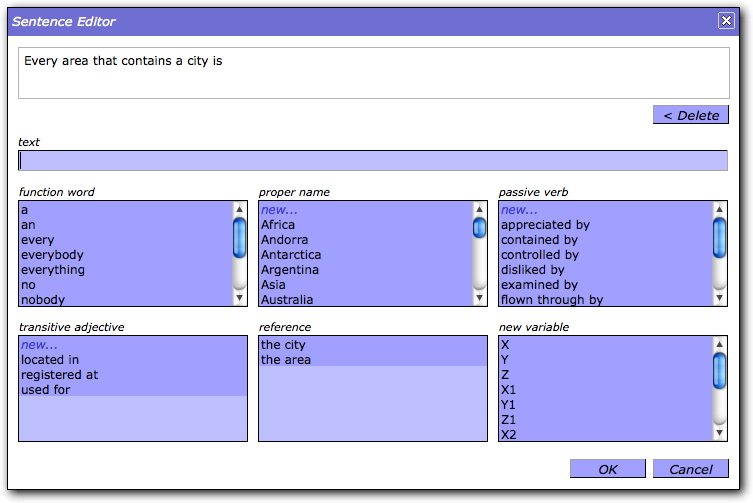}
    \caption{A screenshot of the predictive editor of AceWiki. The fragment ``Every area is'' has already been entered and now the editor shows all possibilities to continue the sentence.}
    \label{fig:editor}
  \end{center}
\end{figure}

In AceWiki, the ontological entities are represented by natural language words and phrases. Proper names (e.g. ``Zurich'', ``Switzerland'', ``Europe'') are interpreted as individuals, nouns (e.g. ``city'', ``country'') are interpreted as classes, and transitive verbs (e.g. ``borders''), \emph{of}-constructs (e.g. ``part of''), and transitive adjectives (e.g. ``located-in'') are interpreted as binary relations. Using those words together with the predefined function words of ACE (e.g.\ ``a'', ``every'', ``if'', ``then'', ``and'', ``not'', ``is'', ``that''), ontological statements are expressed as ACE sentences:
\medskip\\\hspace*{5pt}
  \includegraphics[width=7.8cm]{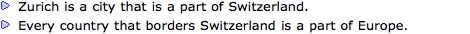}
\smallskip\\
As those examples show, the formal statements are easily readable and understandable by any English speaking person. In order to enable easy creation and modification of ACE sentences, AceWiki integrates a predictive editor that shows step-by-step the words that are syntactically possible at a given position in the sentence. Figure \ref{fig:editor} shows a screenshot of this editor.

\begin{figure}[tb]
  \begin{center}
    \includegraphics[width=8.4cm]{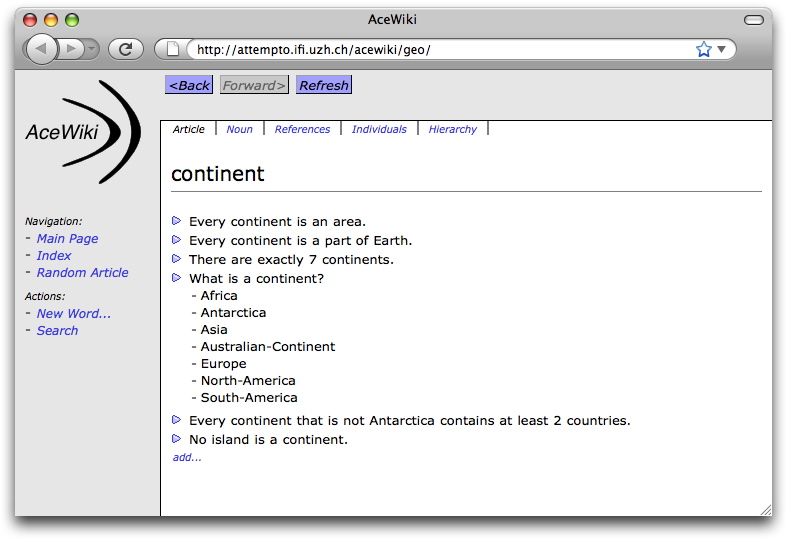}
    \caption{A screenshot of the web interface of AceWiki showing the wiki article for the class ``continent''.}
    \label{fig:window}
  \end{center}
\end{figure}

Each of the ontological entities gets its own wiki article. Figure \ref{fig:window} shows an example. Every article consist of ACE sentences most of which can be translated into OWL, e.g:
\medskip\\\hspace*{5pt}
  \includegraphics[width=7.8cm]{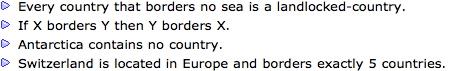}
\smallskip\\
ACE is more expressive than OWL, and thus we can write statements that go beyond the semantic expressivity of OWL (e.g.\ rule-like statements). Such statements are marked with a red triangle (and are currently ignored by the reasoner):
\medskip\\\hspace*{5pt}
  \includegraphics[width=7.8cm]{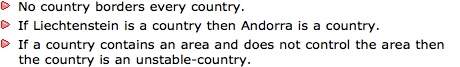}
\smallskip\\
Furthermore, questions can be used to query the knowledge base, e.g:
\medskip\\\hspace*{5pt}
  \includegraphics[width=7.8cm]{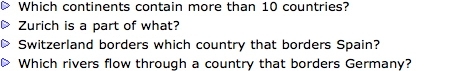}
\smallskip\\
Thus, ACE is an ontology language, a rule language, and a query language at the same time.

AceWiki uses the OWL reasoner Pellet\footnote{\texttt{http://pellet.owldl.com/}} to perform reasoning tasks over the sentences of the wiki that are OWL-compliant. In order to ensure the consistency of the ontology, every new sentence is checked --- immediately after its creation --- whether it would introduce a contradiction. If this is the case then the sentence is not included in the ontology and displayed in red font:
\medskip\\\hspace*{5pt}
  \includegraphics[width=7.8cm]{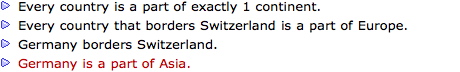}
\smallskip\\
The reasoner is also used to infer the class memberships of individuals. The results are presented in ACE again:
\medskip\\\hspace*{5pt}
  \includegraphics[width=7.8cm]{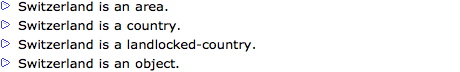}
\smallskip\\
The same is done for class hierarchies:
\medskip\\\hspace*{5pt}
  \includegraphics[width=7.8cm]{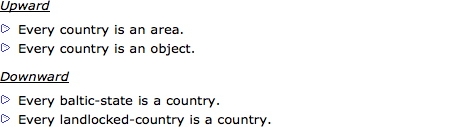}
\smallskip\\
This shows that not only asserted but also inferred knowledge is represented in ACE. Finally, the reasoner is also used to answer questions:
\medskip\\\hspace*{5pt}
  \includegraphics[width=7.8cm]{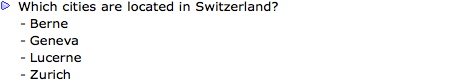}
\smallskip\\
In general, we can say that AceWiki communicates with the users on a very natural level. No knowledge about formal languages is required to deal with AceWiki.

\section{Evaluation}

In our previous work \cite{kuhn08acewiki}, we conducted a user experiment that showed that ordinary people with no background in logic and ontologies are able to deal with AceWiki. The participants --- without being instructed how to interact with the interface --- were asked to add knowledge to AceWiki. About 80\% of the created sentences were correct and sensible. Remarkably, more than 60\% of those sentences were complex in the sense that they contained an implication or a negation.

\section{Conclusions}

AceWiki shows how ontologies can be created and modified in a natural way within a wiki. It demonstrates how semantic wikis using controlled natural language can be expressive and easy to use at the same time. Our previous evaluation showed that AceWiki is indeed easy to learn.

\end{document}